\documentclass[journal,10pt,twocolumn,a4]{IEEEtran}
\usepackage{cite}
\usepackage{amsmath,amssymb,amsfonts}
\usepackage{algorithmic}
\usepackage{graphicx}
\usepackage{textcomp}
\usepackage{mathrsfs}
\usepackage{comment}

\usepackage{enumitem}
\usepackage{color}
\usepackage{multirow}
\usepackage{booktabs}

\begin{document}
\title{Graph Autoencoders for Embedding Learning in Brain Networks and Major Depressive Disorder Identification}

\author{Fuad Noman, Chee-Ming Ting, Hakmook Kang, Rapha\"{e}l C.-W. Phan, Brian D. Boyd, Warren D. Taylor, and Hernando Ombao \vspace{-0.4in}
\thanks{F. Noman, C.-M. Ting, and R. CW Phan are with the School of Information Technology, Monash University Malaysia, Bandar Sunway, Selangor, 47500 Malaysia (e-mail: fuad.noman@monash.edu; ting.cheeming@monash.edu; raphael.phan@monash.edu). }
\thanks{H. Kang is with the Department of Biostatistics, Vanderbilt University Medical Center, Nashville, Tennessee 37232 USA (e-mail: h.kang@Vanderbilt.edu).}
\thanks{W. D. Taylor and B. D. Boyd are with the Center for Cognitive Medicine, Department of Psychiatry, Vanderbilt University Medical Center, Nashville, Tennessee 37212, USA (e-mail: warren.d.taylor@vumc.org; brian.d.boyd@vumc.org).}
\thanks{H. Ombao, is with Statistics Program, King Abdullah University of Science and Technology, Thuwal, 23955-6900 Saudi Arabia. (e-mail: hernando.ombao@kaust.edu.sa).}}

\maketitle

\begin{abstract}
Brain functional connectivity (FC) reveals biomarkers for identification of various neuropsychiatric disorders. Recent application of deep neural networks (DNNs) to connectome-based classification mostly relies on traditional convolutional neural networks using input connectivity matrices on a regular Euclidean grid.
We propose a graph deep learning framework to incorporate the non-Euclidean information about graph structure for classifying functional magnetic resonance imaging (fMRI)-derived brain networks in major depressive disorder (MDD).
We design a novel graph autoencoder (GAE) architecture based on the graph convolutional networks (GCNs) to embed the topological structure and node content of large-sized fMRI networks into low-dimensional latent representations.
In network construction, we employ the Ledoit-Wolf (LDW) shrinkage method to estimate the high-dimensional FC metrics efficiently from fMRI data. We consider both supervised and unsupervised approaches for the graph embedding learning. The learned embeddings are then used as feature inputs for a deep fully-connected neural network (FCNN) to discriminate MDD from healthy controls.
Evaluated on two resting-state fMRI (rs-fMRI) MDD datasets, results show that the proposed GAE-FCNN model significantly outperforms several state-of-the-art methods for brain connectome classification, achieving the best accuracy using the LDW-FC edges as node features. 
The graph embeddings of fMRI FC networks learned by the GAE also reveal apparent group differences between MDD and HC. Our new framework demonstrates feasibility of learning graph embeddings on brain networks to provide discriminative information for diagnosis of brain disorders. 
\end{abstract}

\vspace{-0.05in}

\section{Introduction}


\label{sec:introduction}
\IEEEPARstart{A}{analysis} of brain functional connectivity (FC) networks inferred from functional magnetic resonance imaging (fMRI) data has become an important method to probe large-scale functional organization of the human brain in health and disease \cite{bassett2009}. 
Considerable evidence from rs-fMRI studies have shown altered or aberrant brain functional connectome in various neuropsychiatric and neurodegenerative disorders \cite{woodward2015}, e.g., schizophrenia \cite{venkataraman2012}, autism spectrum disorder (ASD) \cite{muller2011}, Alzheimer's disease (AD) \cite{zhang2010}, suggesting potential use of network-based biomarkers for clinical diagnostics \cite{hallett2020}. Functional abnormalities are detected not only in the strengths of individual connections but also topological structure of resting-state FC networks \cite{bassett2009}. The brain function in major depressive disorder (MDD) --- the most prevalent psychiatric disorder with pervasive depressed mode, cognitive inability and suicidal tendency, has been a subject of intensive studies recently. It is increasingly understood as a network-based disorder with consistent alternations in FC patterns \cite{mulders2015}.
Disrupted resting-state FC from fMRI has been found in MDD core networks, such as the default mode network (DMN) related to self-referential processing and emotion regulation, central executive network (CEN) for attention and working memory, and other subcortical circuitries \cite{brakowski2017}. Increased connectivity within DMN \cite{greicius2007} and decreased connectivity between DMN and CEN have been observed in MDD patients compared to healthy controls (HCs) \cite{mulders2015}. Graph theoretical analyses of rs-fMRI also revealed altered network topological properties in MDD, e.g., enhanced global efficiency \cite{zhang2011} and high local efficiency and modularity \cite{ye2015}.

Machine learning techniques have been increasingly used in turning altered brain FC into biomarkers for fast and automated classification of brain disorders \cite{woo2017}. Vast majority of studies use traditional machine learning algorithms for classification, such as support vector machine (SVM), logistic regression and linear discriminant analysis (For review see \cite{brown2016machine,du2018,dadi2019}). Compared to other disorders, functional connectome-based classification of MDD is relatively unexplored. Several recent studies \cite{zeng2012,cao2014,bhaumik2017multivariate,geng2018,zhu2020cross} have employed SVMs combined with some ad-hoc feature selection methods to differentiate MDD from HCs using rs-fMRI FC, and obtained reasonable classification accuracies on leave-one-subject-out cross-validation. 

Deep learning methods have received significant interest in fMRI-based classification of brain disorders \cite{plis2014}. In recent applications to connectome-based classification, it has shown great potential providing substantial gain in performance over traditional classifiers. Deep neural networks (DNNs) can automatically learn a hierarchy of representations directly from the connectome data, without relying on preliminary feature hand-crafting and selection. Fully-connected DNNs have been used as autoencoders (AE) to map high-dimensional input vectors of FC metrics to latent compact representations for rs-fMRI classification of ASD \cite{heinsfeld2018identification, rakic2020improving} and schizophrenia \cite{kim2016}. Inspired by remarkable success in image and object classification, deep convolutional neural networks (CNNs) have also been used to learn spatial maps of brain functional networks. A CNN architecture (BrainNetCNN) with specially-designed convolutional filters for modeling connectome data was introduced by \cite{kawahara2017brainnetcnn} for predicting neurodevelopment in infants. Various variants of connectome CNNs were subsequently proposed for FC classification. These include one-dimensional (1D) spatial convolutional filters on rs-fMRI FC data for mild cognitive impairment (MCI) identification \cite{meszlenyi2017resting}, 2D-CNNs for FC matrices for ASD classification \cite{sherkatghanad2020automated}, 3D CNNs to combine static and dynamic FC for early MCI detection \cite{kam2019deep}, and multi-domain connectome CNN to integrate different brain network measures \cite{phang2019}. The above-mentioned deep learning models generally neglect the topological information of the brain networks which may lead to sub-optimal performance in brain disorder identification. The flattening of input FC maps in fully-connected DNNs destroys the spatial structure, while the use of fixed 1D or 2D regular grid convolution operators in CNNs also fails to capture the graph-structured connectome data. Brain networks typically exhibit irregular structure with nodes being unordered and connected to a different number of neighbors, which renders convolution operations for regular grid inappropriate for modeling graphs.

Extending deep learning approaches to data in non-Euclidean domain, including graphs, is a rapidly growing field \cite{wu2021}. One popular graph-based neural network (GNN) architecture, the graph convolutional networks (GCNs), generalizes operations in CNNs to learn local and global structural patterns in irregular graphs. A spectral-based GCN has been proposed to perform convolutions in the graph spatial domain as multiplications in the graph spectral domain \cite{bruna2013,kipf2016semi}.
Applications of spectral GCNs to brain disorder detection from brain functional networks are introduced only recently and in its very early stage, e.g., for predicting ASD and conversion from MCI to AD \cite{parisot2017spectral,parisot2018disease,li2019graph,jiang2020hi}. 
These studies used a population graph as input to GCN, where nodes represent subjects with associated resting-state FC feature vectors, while phenotype information is encoded as graph edge weights. However, this approach inherently relies on non-imaging data to construct graphs and requires prior knowledge of relevant phenotype information for specific disorders. Moreover, it is semi-supervised learning using all subjects (both training and testing sets) as inputs and thus lacks generalization on unseen subjects. A recent benchmarking study \cite{he2020} also showed that population-based spectral GCN is less effective than the BrainNetCNN in resting-state FC-based behavioral prediction.

In this paper, we propose a novel framework based on deep GNN for graph embedding on brain functional networks for classifying neuropsychiatric disorders associated with functional dysconnectivity. 
Precisely, we develop a graph autoencoder (GAE) architecture that leverages GCN to encode the non-Euclidean information about brain connectome into low-dimensional latent representations (or network embeddings), on which a decoder is trained to reconstruct the graph structure.
The learned embeddings allow dimensionality reduction of large-sized brain network data, and preserve both the network topological structure and node content information as discriminative features to enhance subsequent connectome-based classification.
The extracted patterns by the multiple graph convolutional layers in GCNs can include high-level representations of nodes' local graph neighborhood.
We utilize the GAE in an inductive framework of embedding generation for network-level classification. In contrast to the GCN used in transductive settings in existing GAEs for a single fixed graph \cite{kipf2016semi,kipf2016variational}, our GAE is designed to generate node embeddings for completely unseen graphs. By learning an embedding function that shared across networks from different subjects, it allows generalization to multiple brain networks of unseen subjects in the downstream brain network classification.
Besides the unsupervised embedding learning using GAE, we also consider supervised learning where the model makes use of disorder class labels to optimize the embeddings. Finally, a readout layer is added to summarize the node representations of each graph into a graph representation, which is then used as feature inputs to a fully-connected DNN (FCNN) for network classification. We apply the proposed GAE-FCNN to rs-fMRI data for classification of MDD and HCs using whole-brain FC networks. The GAE-FCNN is trained on high-dimensional functional networks constructed from rs-fMRI using Ledoit-Wolf (LDW) covariance estimator \cite{ledoit2004well}. We also explore different types of node features: fMRI time series, associated FC edges and local graph measures.
The main contributions of this work are summarized as follows:
\begin{enumerate}[]
\item We propose, for the first time, a graph deep learning framework for brain FC-based identification of MDD.
\item The proposed GAE-FCNN framework offers a novel approach to directly leverage on the alterations in network structure for brain disorder classification via the learned network embeddings. The GCN-based GAE architecture provides a purely unsupervised way to learn embeddings that encode the irregular topological structure of brain networks, which are inadequately modeled by the connectome CNNs and the vectorized FC features in population graphs. The GAE combined with a deep DNN facilitates graph-level classification to predict class labels for the entire brain graph, rather than node/subject-level classification based on population graphs.
\item We demonstrate that our approach outperforms both the BrainNetCNN and population-based GCN by a large margin in identifying MDD based on resting-state functional brain networks from fMRI.
\item We show that high-order network reconstructed from nodes embeddings learned by the proposed GCN-based GAE can reveal differences in network organization between MDD and HCs related to emotion processing.
\end{enumerate}

\begin{figure*}[!t]
\centerline{\includegraphics[width=1.9\columnwidth]{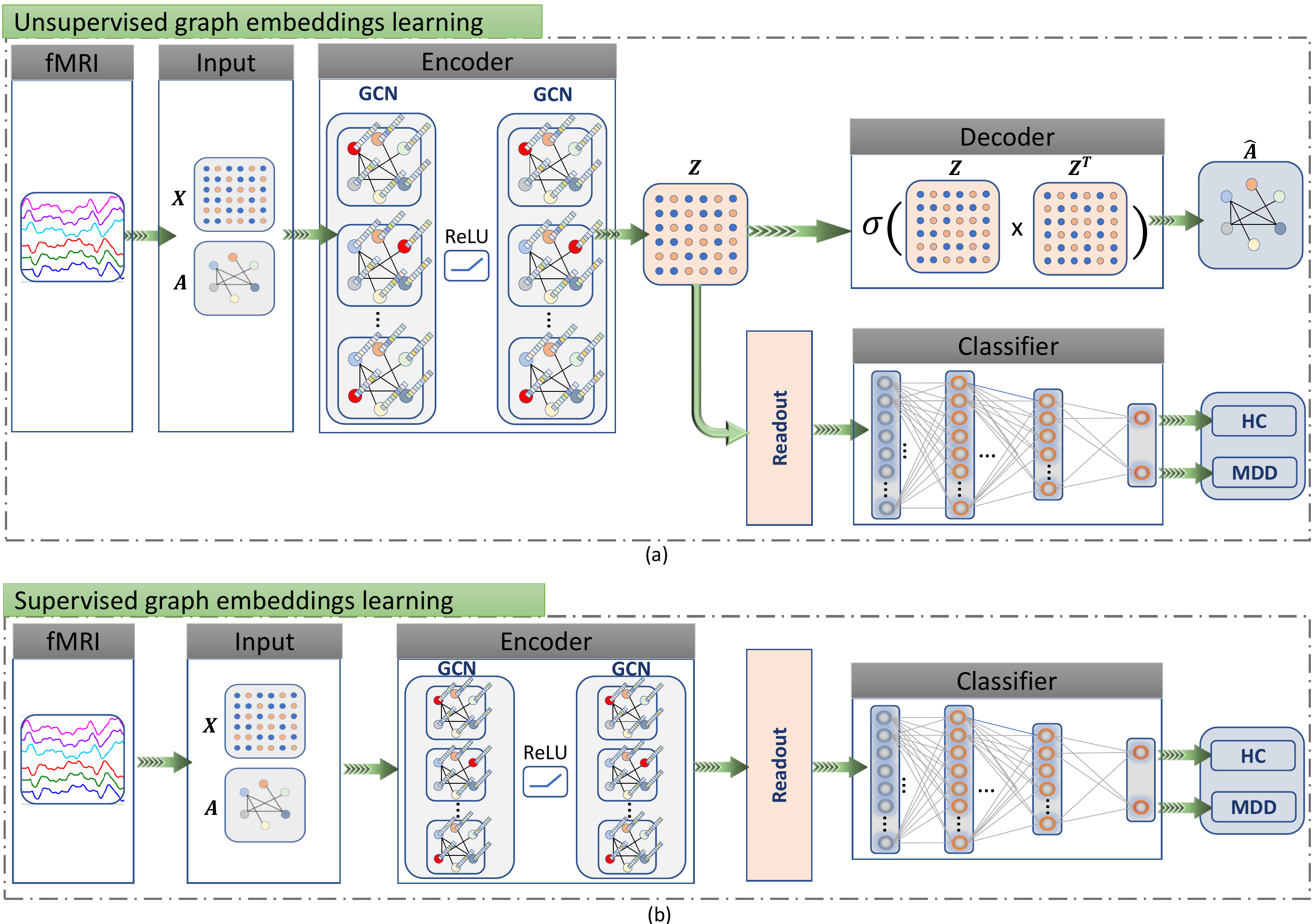}}
\caption{The architecture of proposed GAE-FCNN framework for functional brain network classification. (a) Unsupervised model. The model consists of two components: A GAE employs a GCN-based encoder to encode fMRI connectome data (graph structure $\mathbf{A}$ \& node content $\mathbf{X}$) into latent representations $\mathbf{Z}$ on which a decoder is used to reconstruct the graph information. A deep FCNN performs network-level classification to discriminate MDD patients and HCs based on the learned representations. (b) Supervised model. The GCN encoder combined with FCNN leverages on class labels to learn network representations and performs network classification in an end-to-end framework.}
\label{fig:classifier1_7}
\end{figure*}

\section{rs-fMRI Dataset for MDD}
\subsubsection{Subjects \& Data Acquisition} 
We used a rs-fMRI MDD dataset collected at the Duke University Medical Center, USA, studied previously in \cite{albert2019brain,wang2020}. The Duke-MDD dataset consists of 43 subjects, including 23 non-depressed (HC) and 20 depressed (MDD) participants aged between 20 and 50 years old. Depressed participants had met the Diagnostic and Statistical Manual of Mental Disorder (DSM-IV) criteria of MDD, as assessed by the Mini-International Neuropsychiatric Interview (MINI, version 5.0) \cite{sheehan1998mini} and interview with a study psychiatrist. Participants were scanned on a Siemens 3.0T Trio Tim scanner, with an 8-channel head coil. Echoplanar blood-oxygen-level-dependent (BOLD) functional resting scans were acquired with transverse orientation (TR/TE = 2000/27 ms, voxel size = 4.0 × 4.0 × 4.0 mm, 32 axial slices). A time series of 150 volumes were collected for each scan.\\

\subsubsection{Preprocessing} Standard preprocessing steps were applied to the fMRI data using Conn toolbox (version 15.g) in SPM 12, including motion correction, slice timing correction, co-registration of functional and anatomical images, normalization to the standard MNI template, and Gaussian spatial smoothing with FWHM (full width at half maximum) = 6 mm. The fMRI data were band-pass filtered between 0.01-0.07 Hz. The automated anatomical labeling (AAL) atlas was used to obtain an anatomical parcellation of the whole-brain into 116 regions of interest (ROIs), and ROI-wise fMRI time series were extracted by averaging over voxels.


\section{Methods}

Fig.~\ref{fig:classifier1_7} shows an overview of the proposed GAE-FCNN framework for identifying brain disorders using fMRI-based FC networks, which consists of three stages: (1) Network construction. High-dimensional FC networks are constructed from fMRI data using LDW shrinkage covariance estimator, and associated node features are extracted. (2) Network embedding via a GCN-based GAE. 
The GAE learns network embeddings by using an encoder of stacked GCNs to map the input graph structure and node content of FC networks into latent representation (or embeddings), and using an inner-product decoder to enforce embeddings to preserve graph topological information. (3) Network classification. The learned network embeddings are then used as inputs to a fully-connected DNN to discriminate between MDD patients and HCs. We develop an unsupervised (Fig.~\ref{fig:classifier1_7}(a)) and a supervised (Fig.~\ref{fig:classifier1_7}(b)) framework for learning graph embeddings in brain networks.

\vspace{-0.05in}
\subsection{Connectivity Network Construction}
We consider an undirected graph of brain functional network for each subject, represented by $G \equiv \lbrace V, E\rbrace$ where $V \equiv \lbrace v_1,\ldots, v_N \rbrace$ is a set of $N$ nodes (voxels or ROIs) and $e_{ij} \in E$ denotes the connectivity edge $(i,j)$ between nodes $v_i$ and $v_j$.
The topological structure of the graph $G$ can be represented by an adjacency matrix $\mathbf{A} = [a_{ij}] \in {\{0,1\}}^{N \times N}$, where $a_{ij}=1$ if nodes $v_i$ and $v_j$ are connected, otherwise $a_{ij}=0$. We denote by $\mathbf{X} = [\mathbf{x}_1, \ldots, \mathbf{x}_N]^T \in \mathbb{R}^{N \times d}$ the node feature matrix for $G$, with $\mathbf{x}_i \in \mathbb{R}^d$ representing the content feature vector associated with each node $v_i$.
\subsubsection{Network Connectivity}
In constructing FC networks, we compute the FC matrix based on the temporal correlations of fMRI time series between pairs of ROIs. Let $\mathbf{y}_t \in \mathbb{R}^N$, $t=1, \ldots, T$ be the fMRI time series of length $T$ measured from the $N$ ROIs.
For large-sized fMRI-derived networks in which the number of nodes $N$ is larger or comparable to the number of scans $T$, traditional sample correlation matrix is no longer a reliable and accurate estimator of FC. This is due to large number of correlation coefficients (i.e., $N(N-1)/2$) to be estimated relative to the sample size. This condition applies to the MDD fMRI data considered here ($T = 150$ and $N = 116$ ROIs).
To estimate functional connectomes efficiently, we use the Ledoit-Wolf (LDW) regularized shrinkage estimator \cite{ledoit2004well,brier2015} which can yield well-conditioned FC estimates in high-dimensional settings when the ratio of $N/T$ is large. The LDW covariance estimator is defined by $\widetilde{\boldsymbol{\Sigma}}=(1-\alpha) \widehat{\boldsymbol{\Sigma}}+\alpha \boldsymbol{\Delta}$ with
$\boldsymbol{\Delta}=(Tr(\widehat{\boldsymbol{\Sigma}})/N)\mathbf{I}_N$, where
$\alpha$ is a shrinkage parameter, $\mathbf{I}_N$ is a $N \times N$ identity matrix, $\operatorname{Tr}(\cdot)$ is the trace and 
$\widehat{\boldsymbol{\Sigma}}=\frac{1}{T}\sum_{t=1}^{T}(\mathbf{y}_t-\bar{\mathbf{y}}) (\mathbf{y}_t-\bar{\mathbf{y}})^{T}$ is the $N \times N$ sample covariance matrix with sample mean $\bar{\mathbf{y}} = \frac{1}{T} \sum_{t=1}^{T} \mathbf{y}_t$.
The shrinkage coefficient $\alpha$ can be estimated data-adaptively \cite{chen2010shrinkage}.
The correlation matrix is then computed as 
$\mathbf{R}=\mathbf{D}^{-1/2} \widetilde{\boldsymbol{\Sigma}} \mathbf{D}^{-1/2}$ where $\mathbf{D} = diag(\widetilde{\boldsymbol{\Sigma}})$.

We can generate the adjacency matrix $\mathbf{A}$ by thresholding the correlation matrix $\mathbf{R}$. We used the proportional thresholding \cite{van2017proportional} which sets a proportion $\tau$ of strongest connections (with the highest absolute correlation values) of the derived FC matrix for each individual network to 1, and other connections to zero. By applying a proportional threshold value of $\tau$, the number of retained links/edges in a graph is $\tau (N^2-N)/2$. This approach will result in a fixed density of edges in graphs across all subjects, and thus enabling meaningful comparison of network topology between different groups and conditions. It can also generate more stable network metrics compared to the absolute thresholding \cite{garrison2015}. It has been shown that the setting of threshold $\tau$ has a significant impact on the overall performance of the network classification model \cite{jiang2020hi}. Besides, when $\tau$ decreases, networks become sparser and may lead to the zero-degree nodes (isolated nodes totally disconnected from the rest of the graph). By evaluating over a range of thresholds, $\tau=0.4$ was chosen to generate graphs without zero-degree nodes for all subjects and give the optimal classification performance based on the validation set.

\subsubsection{Node Features} We consider three types of node features for $\mathbf{X}$. (1) ${T \times 1}$ raw rs-fMRI time series associated with each node which can capture spontaneous fluctuations in the BOLD signal in individual brain regions. (2) ${N \times 1}$ FC weights of edges connected to each node, i.e., each column of the LDW-estimated correlation matrix. (3) Graph-theoretic measures to characterize graph topological attributes at local (nodal) level. A list of 18 different nodal graph measures \cite{rubinov2010complex} was extracted for each individual node, including degree, eigenvector centrality, modularity, PageRank centrality, nodal eccentricity, community Louvain, module degree z-score, participation coefficient, routing efficiency, clustering coefficient, diversity coefficient, gateway coefficient (node strength), gateway coefficient (betweenness centrality), local assortativity, participation coefficient, node strength, node betweenness, and global efficiency.

\vspace{-0.05in}
\subsection{Graph Convolutional Autoencoder}
We propose a new approach that builds on the graph autoencoder (GAE) \cite{kipf2016variational,pan2018} to learn graph embeddings on brain networks in a purely unsupervised framework. Given the brain network $G$ for each subject, the autoencoder maps the nodes $v_i \in V$ to low-dimensional vectors $\mathbf{z}_i \in \mathbb{R}^k$ (or embeddings), using an \textit{encoder} $f:(\mathbf{A},\mathbf{X}) \mapsto \mathbf{Z}$ where $\mathbf{Z} = [\mathbf{z}_1,\ldots,\mathbf{z}_N]^T \in \mathbb{R}^{N \times k}$ with $k<<N$ the dimension of embedding, and then reconstruct the graph structure from the embeddings $\mathbf{Z}$ using a \textit{decoder}. The learned latent representations $\mathbf{Z}$ should reflect the topological structure of the graph $\mathbf{A}$ and the node content information $\mathbf{X}$. It contains all the information necessary for downstream graph classification tasks for brain disorders. We consider two variants of GAE: (1) Generic GAE which aims to reconstruct the original input graph adjacency matrix, (2) Variational GAE (VGAE) \cite{kipf2016variational}, a variational extension of GAE to learn the distribution of embeddings, which could prevent potential model overfitting. The GAE proposed originally in \cite{kipf2016variational} was applied for \textit{transductive} problems, e.g., to make semi-supervised node or link prediction within a single fixed graph. In contrast, we apply the GAE in an \textit{inductive} setting for multi-graph representation learning for whole-network classification, where our GAE is trained on multi-subject brain networks from the training set, and the trained graph encoder is then used to generate embeddings for completely unseen networks in the test set for subsequent classification. The weight parameters of our graph encoder are shared among networks of different subjects, which allows learning of graph representations across subjects and generalization over unseen graphs.

\subsubsection{Graph Convolutional Encoder Model}

To encode both graph structure $\mathbf{A}$ and node content $\mathbf{X}$ into $\mathbf{Z}$ in a unified way, we employ a variant of graph convolutional network (GCN) \cite{kipf2016semi} as the graph encoder of GAE. The GCN is a first-order approximation of graph convolutions in the spectral domain.
The multi-layer GCN learns a layer-wise transformation by a spectral graph convolutional function $f$
\begin{equation}\label{Eqn:1}
\mathbf{Z}^{(l+1)} = f(\mathbf{Z}^{(l)},\mathbf{A}|\mathbf{W}^{(l)})
\end{equation}
where $\mathbf{Z}^{(l)}$ is the latent feature matrix after convolution at $l$-th layer of GCN with layer-dependent dimensions, $\mathbf{W}^{(l)}$ is a layer-specific trainable weight matrix. Here, $\mathbf{Z}^{(0)} = \mathbf{X} \in \mathbb{R}^{N \times d}$ is the input node feature matrix. The propagation for each layer of the GCN can be calculated as
\begin{equation}
\mathbf{Z}^{(l+1)} = \sigma\left(\tilde{\mathbf{D}}^{-\frac{1}{2}} \tilde{\mathbf{A}} \tilde{\mathbf{D}}^{-\frac{1}{2}} \mathbf{Z}^{(l)} \mathbf{W}^{(l)}\right)
\label{Eqn:2}
\end{equation}
where $\tilde{\mathbf{A}} = \mathbf{A} + \mathbf{I}_N$ is normalized adjacency matrix with added self-connections to ensure numerical stability, $\tilde{\mathbf{D}}$ is a node degree matrix with diagonals $\tilde{d}_{ii}=\sum_{j} (\tilde{a}_{ij})$, and $\sigma(.)$ denotes the activation function. Model (\ref{Eqn:2}) generates embeddings for a node by aggregating feature information from its local neighborhood at each layer.

We construct the graph convolutional encoder based on two-layered GCN as in \cite{pan2018}
\begin{align}
\mathbf{Z}^{(1)} = & f_{\text{relu}}(\mathbf{X},\mathbf{A}|\mathbf{W}^{(0)}) \label{eq:en1} \\
\mathbf{Z}^{(2)} = & f_{\text{linear}}(\mathbf{Z}^{(1)},\mathbf{A}|\mathbf{W}^{(1)}) \label{eq:en2}
\end{align}
which produces latent representation $\mathbf{Z}$ with the following forward propagation
\begin{equation}
\mathbf{Z} = GCN(\mathbf{A},\mathbf{X}) = \sigma_{1}\left(\bar{\mathbf{A}}\sigma_{0}\left(\bar{\mathbf{A}}\mathbf{X} \mathbf{W}^{(0)}\right)\mathbf{W}^{(1)}\right)         
\label{Eqn:3}
\end{equation}
where $\bar{\mathbf{A}} = \tilde{\mathbf{D}}^{-\frac{1}{2}} \tilde{\mathbf{A}} \tilde{\mathbf{D}}^{-\frac{1}{2}}$, $\sigma_0$ and $\sigma_1$ are ReLU(·) and linear activation functions in first and second layers, respectively.

In the VGAE, variational graph encoder is defined by an inference model parameterized by a two-layer GCN \cite{kipf2016variational}
\begin{align}
 q(\mathbf{Z}|\mathbf{X},\mathbf{A}) = & \prod_{i=1}^N  q(\mathbf{z}_i|\mathbf{X},\mathbf{A}), \label{eq:vae1} \\
 q(\mathbf{z}_i|\mathbf{X},\mathbf{A}) =  & N(\mathbf{z}_i|\boldsymbol{\mu}_i,\text{diag}(\boldsymbol{\sigma}_i^2)) \label{eq:vae2}
\end{align}
Here, the embeddings $\mathbf{z}_i$ are generated according to a normal distribution $q(.)$ with mean $\boldsymbol{\mu}_i$ and variance $\boldsymbol{\sigma}_i^2$. $\boldsymbol{\mu} = GCN_{\boldsymbol{\mu}}(\mathbf{A},\mathbf{X})$ is the matrix of mean vectors $\boldsymbol{\mu}_i$ defined by the GCN encoder output in (\ref{Eqn:3}), and $\text{log}\boldsymbol{\sigma} = GCN_{\boldsymbol{\sigma}}(\mathbf{A},\mathbf{X})$ is defined similarly for $\boldsymbol{\sigma}_i^2$ using another encoder output. 

\subsubsection{Decoder Model} The decoder of GAE aims to decode graph structural information from the embeddings by reconstructing the graph adjacency matrix. The GAE decoder model predicts the presence of a link between two nodes for the input graph $\mathbf{A}$ based on the inner-product between latent vectors of $\mathbf{Z}$:
\begin{equation}
\begin{aligned}
p(\mathbf{A} \mid \mathbf{Z})= & \prod_{i=1}^{N} \prod_{j=1}^{N} p\left({a}_{ij} \mid \mathbf{z}_{i}, \mathbf{z}_{j}\right), \\
& \text {with } p\left(a_{ij}=1 \mid \mathbf{z}_{i}, \mathbf{z}_{j}\right)=\sigma\left(\mathbf{z}_{i} \mathbf{z}_{j}^{\top}\right)
\end{aligned}
\label{Eqn:dec}
\end{equation}
\noindent where $\sigma(.)$ is the logistic sigmoid function. The graph adjacency matrix can be reconstructed as $\hat{\mathbf{A}} = \sigma(\mathbf{Z}\mathbf{Z}^T)$.


\subsubsection{Optimization} Given a dataset of brain networks of $R$ subjects $\{G^1, \ldots, G^R\}$ where each network $G^r$ is attributed with $(\mathbf{X}^{r},\mathbf{A}^{r})$. The GAE is trained by maximizing the expected negative reconstruction error of the graphs over all subjects in the dataset
\begin{equation}
\begin{aligned}
\mathcal{L}\left(\{\mathbf{X}^{r},\mathbf{A}^{r}\}_{r=1}^R\right) & = \sum_{r=1}^{R} \mathcal{L}(\mathbf{X}^{r},\mathbf{A}^{r}) \\
\text{with} \ \mathcal{L}(\mathbf{X}^{r},\mathbf{A}^{r}) & = \mathbb{E}_{q(\mathbf{Z} \mid(\mathbf{X}^{r}, \mathbf{A}^{r}))}[\log p(\mathbf{A}^{r} \mid \mathbf{Z})]. \label{Eqn:gaeL}
\end{aligned}
\end{equation}
Since the ground-truth adjacency matrix $\mathbf{A}$ is sparse, the optimization is constrained by of non-zero elements of $\mathbf{A}$ (i.e., $a_{ij} = 1$).
For the VGAE, we maximize the variational lower bound w.r.t the parameters $\mathbf{W}$
\begin{multline}
\mathcal{L}(\mathbf{X}^{r},\mathbf{A}^{r})=\mathbb{E}_{q(\mathbf{Z} \mid(\mathbf{X}^{r}, \mathbf{A}^{r}))}[\log p(\mathbf{A}^{r} \mid \mathbf{Z})] \\ – KL[q(\mathbf{Z}\mid \mathbf{X}^{r},\mathbf{A}^{r}) \| p(\mathbf{Z})]
\label{Eqn:vgaeL}
\end{multline}
where $KL(.)$ is the Kullback-Leibler divergence function that measures the distance between two distributions. We use a Gaussian prior $p(\mathbf{Z}) = \prod_i p(\mathbf{z}_i) = \prod_i N(\mathbf{z}_i|\mathbf{0},\mathbf{I})$. We perform mini-batch gradient descent and make use of the reparametrization trick \cite{kingma2013auto} for training.

\vspace{-0.05in}
\subsection{GAE-FCNN for Network Classification}

We design a GAE-FCNN framework for brain connectome classification by combining the GAE with a fully-connected DNN (FCNN). A readout layer is added to summarize latent node representations $\mathbf{Z}$ learned by the GAE for each graph into graph-level representations, which are then fed into an FCNN to classify individual networks into MDD and HC.

\subsubsection{Graph Embeddings Vectorization (Readout)}
We apply a readout operation on the network node representations to generate higher graph-level representations. In the readout layer, a vector representation $\mathbf{z}_G \in \mathbb{R}^{k} $ of the graph $G$ can be learned by aggregating all individual node embeddings in the graph via some statistical summary measures
\begin{equation}
\mathbf{z}_G = \text{mean/max/sum}(\mathbf{z}^{(L)}_1,\ldots,\mathbf{z}^{(L)}_N)
\label{Eqn:readout}
\end{equation}
where $L$ is the index of the last graph convolutional layer. The graph embedding $\mathbf{z}_G$ can then be used to make predictions about the entire graph. The mean/max/sum-based embeddings can be used individually or concatenated into a single vector to capture different graph-level information. In addition, to retain embedding information for all nodes, we also compute the graph embedding as $\mathbf{z}_G = \text{vec}(\mathbf{Z})$ by flattening of $\mathbf{Z}$.

\subsubsection{FCNN Classifier} The graph vector embeddings $\mathbf{z}_G$ are then used as inputs to a deep FCNN for network-level classification. The FCNN classifier consists of multiple fully-connected/dense layers, plus a final softmax classification layer to output the predictive probabilities of class labels for each network. The dense layer approximates a non-linear mapping function to further capture relational information in the graph embeddings to discriminate between MDD and HC. The weight parameters of the FCNN are trained by minimizing cross-entropy loss function using stochastic gradient descent methods and backpropagation of error. Dropout is also applied to prevent overfitting \cite{srivastava2014dropout}.

\subsubsection{Supervised \& Unsupervised Embedding Learning}
We consider two classification schemes using the network embeddings learned in supervised and unsupervised ways. The proposed encoder-decoder framework (Fig.~\ref{fig:classifier1_7}(a)) to extract network embeddings described thus far is by default unsupervised, i.e., the GAE is trained to reconstruct the original graph structure. The unsupervised learning makes use of only information in $\mathbf{A}$ and $\mathbf{X}$, without knowledge of a particular downstream connectomic classification task. We further develop a supervised framework, as shown in Fig.~\ref{fig:classifier1_7}(b), which utilizes the task-specific classification labels in order to learn the network embeddings. The inner-product decoder in the supervised model is replaced with an FCNN to decode the embeddings from the output of GCN encoder to class labels. The parameters of the GCN encoder can be trained based on cross-entropy loss between the predicted and true class labels using the backpropagation algorithm. By incorporating task-specific supervision, the encoder model is optimized to generate embeddings that may be more discriminative of the MDD and HC classes. This model provides an end-to-end framework for the brain network classification.

\section{Experiments}

In this section, we present experimental evaluation of the proposed GAE-FCNN models for connectome classification on the rs-fMRI MDD dataset described in Section II.

\vspace{-0.05in}
\subsection{Experimental Setup}

\subsubsection{Data Partitioning} 
We applied a nested-stratified 5-fold cross-validation (CV) data partitioning scheme \cite{pereira2009machine} to evaluate the performance of different models in classifying MDD and HC. Specifically, a two-level 5-fold CV was used comprising an outer-loop for testing and an inner-loop for model hyper-parameter optimization. For each iteration in the outer-loop, a test set was assigned, and the rest of the data were split into five train-validation partitions to tune the model hyper-parameters. This process was repeated for all outer-loop 5-fold partitions. The best performing model (on the validation set) of the five candidate models was then selected to evaluate the performance on the unseen test sets.
The classification performance were evaluated using the following metrics: classification accuracy ($Acc$), sensitivity ($Sen$), specificity ($Spe$), precision ($Pre$), and F-score ($F_1$).

\subsubsection{Model Architecture and Training}
\begin{figure}[!t]
\centerline{\includegraphics[width=0.9\columnwidth]{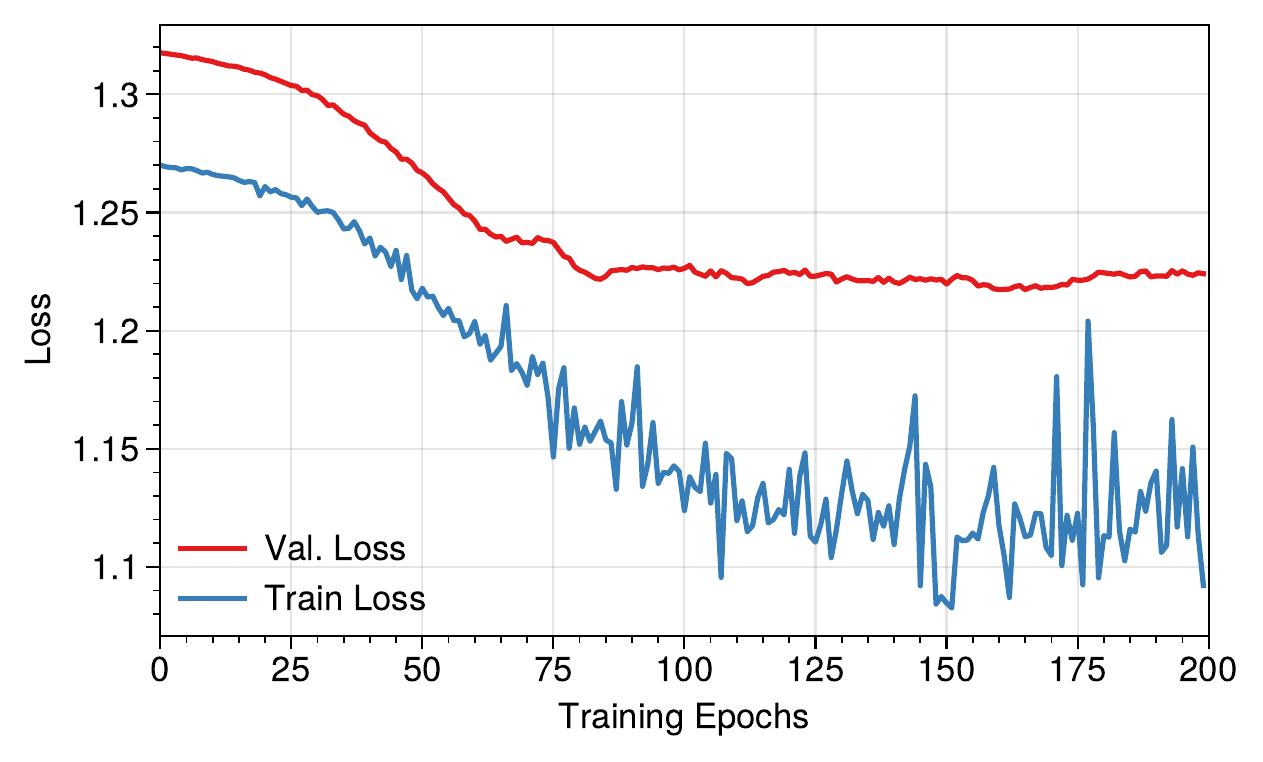}}
\vspace{-0.05in}
\caption{The learning curve as a function of training epochs and reconstruction loss of the GAE model.}
\label{fig:loss_fig}
\vspace{-0.15in}
\end{figure}

We implement the proposed GAE-FCNN based on PyTorch \cite{NEURIPS2019_9015} using the GraphConv module from DGL library \cite{wang2019dgl} for GCN. For the unsupervised model, the architecture and hyper-parameters of GAE and FCNN were determined separately. We computed the reconstruction error of graph over a range of hyper-parameters for the GAE, and a two-layered GCN with respective embedding dimensions of 64 and 16 was identified as the optimal architecture for both GAE/VGAE with the minimum reconstruction error. Further increase in the number of GCN layers gave no further improvement. Using the extracted $116 \times 116$ network adjacency matrices and $116 \times d$ node feature matrices (dimension $d$ depends on type of features used) as inputs, the GAEs were trained using Adam optimizer \cite{AdamOpt} to minimize graph reconstruction loss, with learning rate of $1.00 \times 10^{-4}$, reduce-factor of $0.5$, 200 training epochs and a batch size of 8. Fig.~\ref{fig:loss_fig} illustrates a training curve of the GAE model with decreasing reconstruction error over epochs. The trained GAE decoder was then used to generate $116 \times 16$ node embedding matrices $\mathbf{Z}$ as inputs to the FCNN. Bayesian optimization \cite{skopt} with Expected Improvement (EI) acquisition function was used to optimize the hyper-parameters of FCNN, which suggested an architecture of 3 dense layers (with respective 256, 256, 128 hidden nodes), learning rate of $0.01$, reduce factor of $7.31 \times 10^{-1}$ and a batch size of 4. The FCNN was also trained on the extracted graph embeddings $\mathbf{z}_G$ using Adam algorithm.

For the supervised model, the hyper-parameters of the GCN and FCNN were optimized simultaneously using the Bayesian optimizer. The selected hyper-parameters are: 1 convolutional layer with dimension of 94 for GCN, 2 dense layers (with 128, 64 hidden nodes) for FCNN with learning rate of $1.2 \times 10^{-6}$, reduce factor of $6.13 \times 10^{-1}$ and a batch size of 6. A dropout ratio of 0.2 was also chosen for the dense layers. The model was trained on the fMRI network data with target class labels, using the Adam algorithm to minimize cross-entropy loss.

\subsubsection{Methods for Comparison} We benchmark the classification performance of the proposed methods with traditional SVM classifier and state-of-the-art connectome-specific DNN models: BrainNetCNN and four GCN-based methods. These competing models were evaluated with the same 5-fold CV as the proposed methods.
\begin{enumerate}[]
\item \textit{SVM-RBF}: We trained SVM with radial basis function (RBF) on the vectorized LDW-correlation coefficients. 
\item \textit{BrainNetCNN}: The BrainNetCNN \cite{kawahara2017brainnetcnn} is a specially designed deep CNN model which can preserve spatial information in brain connectivity data. Here, the $N \times N$ LDW correlation matrices were used directly as inputs to the BrainNetCNN to predict the class labels of MDD and HC as output. It consists of three types of layers: edge-to-edge (E2E) layers, edge-to-node (E2N) layers, and node-to-graph (N2G) layers. The E2E layer applies a cross-shaped convolution filter to each element of the FC input matrix, and combines the edge weights of neighbor nodes to output an $N \times N$ matrix. The E2N layer is equivalent to the 1D-CNN filter designed for dimensionality reduction. The N2G layer is a dense layer taking the $N \times 1$ E2N output to produce a single scalar. Finally, the output of N2G is fed to classification layer for prediction.
\item \textit{Population-based GCN}: This method exploits GCN to model a population graph, where each node represents a subject and edges encode similarity between subjects \cite{kipf2016semi}. It performs node/subject level-classification in a semi-supervised manner to predict brain disorders. Similar to \cite{kipf2016semi}, we used the vectorized upper triangular part of LDW correlation matrices as inputs to the population-based GCN. We set the model hyper-parameters with Chebyshev polynomial basis filters for spectral convolutions as in \cite{kipf2016semi}. The model was trained using 500 epochs with early stopping patience of 10 epochs.
\item \textit{GroupINN} \cite{yan2019groupinn}: 
The group-based GCN (GroupINN) uses an ensemble of GCNs to learn graph-level latent embedding representations. The unified framework uses multi-graph clustering and embedding learning to jointly optimize the training process of graph convolutions. 
\item \textit{Hi-GCN} \cite{jiang2020hi}: Hierarchical GCN (Hi-GCN) is a two-level GCN. The first level learns topological embeddings from brain connectivity networks of individual subjects. The second level is a population-based GCN using individual network embedding as node features to incorporate contextual associations between subjects for classification. It can jointly learn the graph embeddings from the brain FC and population networks at the same time. 
\item \textit{E-Hi-GCN} \cite{li2021te}: An ensemble of of Hi-GCN (E-Hi-GCN) is an ensemble framework combining a set of Hi-GCNs each of which is trained on different sparsity level brain networks. It is capable of handling high-dimensional noisy correlations in brain networks. 

\end{enumerate}

We applied hyper-parameter tuning using the Bayesian optimization on both the proposed and competing methods based on the same cross-validation setting to obtain the optimal set of hyper-parameters for each method. The involved hyper-parameters and their search range used in the parameter tuning are given in Appendix Table.~\ref{params}.

\subsection{Results}

\subsubsection{Comparison of Network Construction Strategies}

Table \ref{table-net} shows the classification performance (average and standard deviation over 5 folds) of the unsupervised GAE/VGAE-FCNN and supervised GCN-FCNN classifiers. To investigate the impact of choices of network construction strategies on classification, we also evaluated two FC metrics to construct the graph adjacency matrix $\mathbf{A}$: Pearson's correlation matrix and LDW shrinkage correlation matrix; three types of input node features for $\mathbf{X}$: raw rs-fMRI time series, FC weights (LDW correlation coefficients) and nodal graph-theoretic measures. The selected readout schemes are also given, and details will be discussed in the next section.
As expected, using input graph data based on the LDW correlations shows superior performance over the traditional Pearson's correlations in classifying MDD and HC for all classification models, as the LDW shrinkage method can provide more reliable estimate of the high-dimensional network structure. For node features, the use of LDW-FC generally provided better classification than the raw fMRI time series and local graph measures. This indicates more discriminative information in the connection weights compared to the low-level BOLD fluctuations, and learning of higher-level meta representations from local graph features also fails to offer additional advantages for classification.

\begin{figure}[!t]
\centerline{\includegraphics[width=0.9\columnwidth]{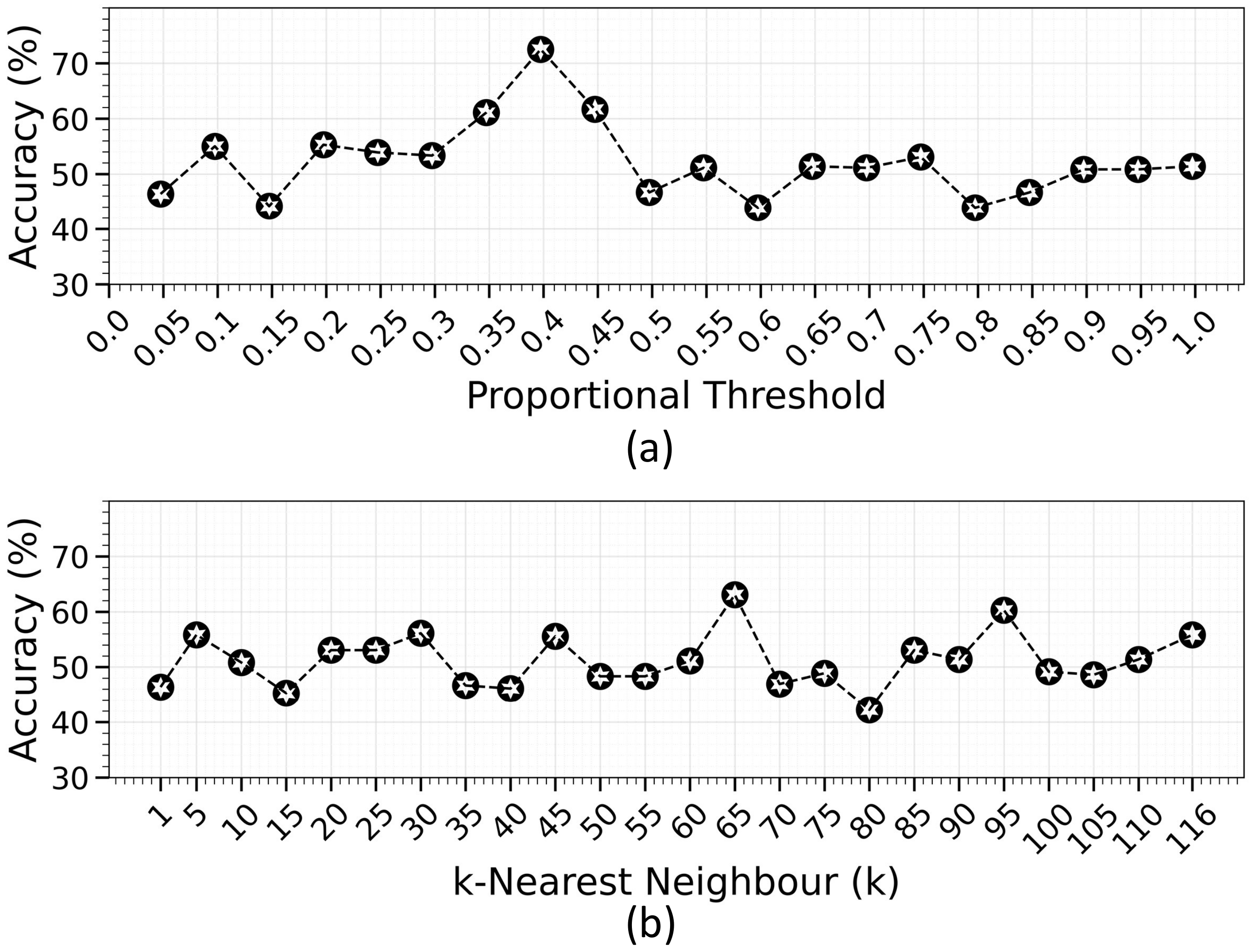}}
\caption{Effect of two FC network thresholding strategies with varying threshold values on MDD classification accuracy of the unsupervised GAE-FCNN model. (a) Proportional thresholding. (b) k-nearest neighbor graph. Network adjacency matrix: LDW. Node feature: LDW-FC.}
\label{fig:PT_KNN}
\vspace{-0.1in}
\end{figure}

\begin{figure}[!t]
\centerline{\includegraphics[width=0.95\columnwidth]{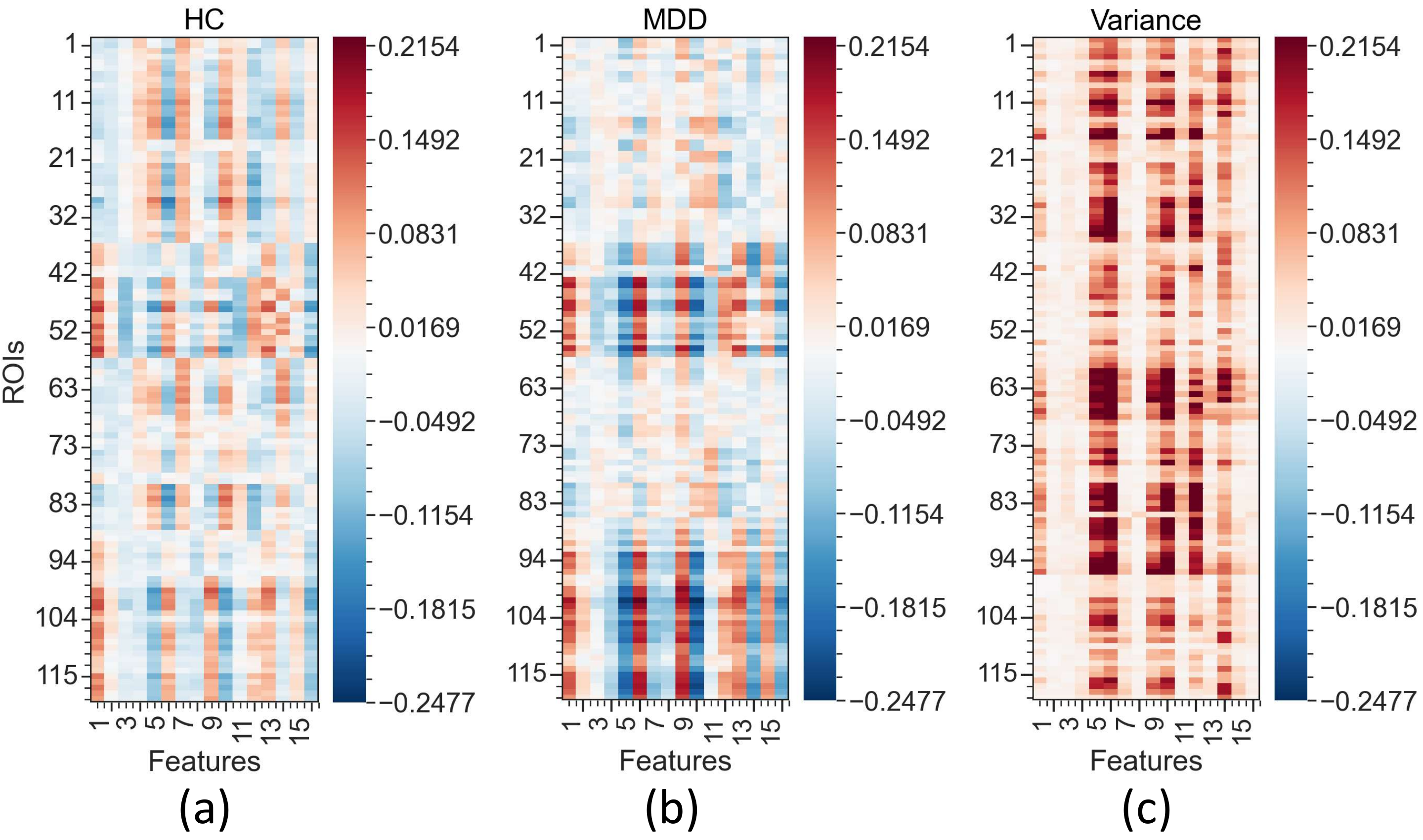}}
\caption{Visualization of $116 \times 16$ network embedding matrices $\mathbf{Z}$ (averaged over subjects) learned by the GCN-GAE from rs-fMRI functional networks. (a) HC subjects. (b) MDD subjects. (c) The between-group variance i.e., $\boldsymbol{\sigma}^{2}_\text{b} = \left(\bar{\mathbf{Z}}_\text{HC}-\bar{\mathbf{Z}}\right)^{2} n_\text{HC} + \left(\bar{\mathbf{Z}}_\text{MDD}-\bar{\mathbf{Z}}\right)^{2} n_\text{MDD}$ where $\bar{\mathbf{Z}}_\text{HC}$ and $\bar{\mathbf{Z}}_\text{MDD}$ are mean embeddings of each group, $\bar{\mathbf{Z}}$ is mean embeddings over both groups, $n_\text{HC}$ and $n_\text{MDD}$ are the number of subjects in each group.}
\vspace{-0.05in}
\label{fig:res_fig1}
\vspace{-0.15in}
\end{figure}

We can see that the unsupervised GAE/VGAE-FCNNs performed better than the supervised GCN-FCNN model, with GAE-FCNN achieving the highest classification accuracy when using LDW-FC for both the graph construction and node features. This suggests that embeddings learned in an unsupervised manner to preserve faithfully the brain network topology can be more predictive of MDD and HC than that optimized to discriminate the class labels directly. Among the unsupervised models, however use of the probabilistic encoding framework in VGAE does not improve classification performance, probably limited by the strong assumption of an \textit{i.i.d.} Gaussian prior on latent embeddings, and the approximated model parameter inference of the variational method. Future work will investigate better-suited prior distribution in the VGAE for brain network data.

\begin{table*}[!t]
\centering
\caption{Classification Performance of the proposed GAE/VGAE-FCNN and supervised GCN-FCNN models using different network construction strategies for classifying MDD and HC subjects based on rs-fMRI functional networks. Networks are constructed based on Pearson's and LDW correlation matrices using proportional thresholds of $\tau$. Results are averages (standard deviations) of performance measures over 5-fold cross-validation.
}
\setlength{\tabcolsep}{5pt}
\renewcommand{\arraystretch}{1.2}
\label{table-net}
\resizebox{1\linewidth}{!}{
\begin{tabular}{lllccccccc}
\hline \hline
Classifier& Adjacency $\mathbf{A}$ & Node Feature $\mathbf{X}$ & Readout & Acc & Sen & Spe & Pre & F1\\
\toprule 
    \multirow{6}{*}{\vtop{\hbox{\strut Unsupervised}\hbox{\strut GAE-FCNN}}}& \multirow{3}{*}{\vtop{\hbox{\strut Pearson}\hbox{\strut ($\tau= $0.25)}}}& Raw-fMRI& [mean,max,sum]& 35.00 $\pm$ \, 2.04&  40.00 $\pm$ 12.25&  31.00 $\pm$ 12.00 & 33.00 $\pm$ \, 4.76&  35.60 $\pm$ \, 7.47\\
    & & Graph-measures& flatten& 57.50 $\pm$ 13.82&  30.00 $\pm$ 29.15&  83.00 $\pm$ 23.58 & 50.00 $\pm$ 44.72&  33.33 $\pm$ 28.60\\
    & & Pearson-FC&[mean,max,sum] &58.06 $\pm$ \, 9.15&	70.00 $\pm$ 18.71&	48.00 $\pm$ 25.02 & 56.67 $\pm$ \, 9.33&	60.31 $\pm$ \, 6.44\\
    \cline{2-9}
    & \multirow{3}{*}{\vtop{\hbox{\strut LDW}\hbox{\strut ($\tau= $0.4)}}}& Raw-fMRI&	[mean,max,sum]& 60.56 $\pm$ \, 4.36&  45.00 $\pm$ 33.17&  74.00 $\pm$ 19.34 & 48.10 $\pm$ 24.85&  44.07 $\pm$ 25.38\\
    & & Graph-measures& [mean,max,sum]& 69.72 $\pm$ \, 9.06&  55.00 $\pm$ 18.71&  83.00 $\pm$ 15.36 & 80.00 $\pm$ 18.71&  61.33 $\pm$ 14.04\\
    & & LDW-FC  &  flatten & \textbf{72.50 $\pm$ 10.77} & \textbf{60.00 $\pm$ 20.00} & \textbf{83.00 $\pm$ 15.36} & \textbf{80.00 $\pm$ 18.71} & \textbf{65.14 $\pm$ 17.20} \\
    
    \hline
    \multirow{6}{*}{\vtop{\hbox{\strut Unsupervised}\hbox{\strut VGAE-FCNN}}}& \multirow{3}{*}{\vtop{\hbox{\strut Pearson}\hbox{\strut ($\tau= $0.25)}}}&Raw-fMRI& flatten& 57.50 $\pm$ 22.08&	60.00 $\pm$ 30.00&	55.00 $\pm$ 21.91 & 52.67 $\pm$ 21.87&	55.49 $\pm$ 25.18\\
    & & Graph-measures& flatten & 58.33 $\pm$ 20.24&  55.00 $\pm$ 33.17&  62.00 $\pm$ 31.08 & 50.17 $\pm$ 28.68&  50.63 $\pm$ 28.68\\
    & & Pearson-FC& flatten & 64.72 $\pm$ \, 8.94&	\textbf{65.00 $\pm$ 12.25}&	64.00 $\pm$ 13.56 & 62.33 $\pm$ \, 8.27&	\textbf{63.10 $\pm$ \, 8.65}\\
    \cline{2-9}
    & \multirow{3}{*}{\vtop{\hbox{\strut LDW}\hbox{\strut ($\tau= $0.4)}}}& Raw-fMRI&	[mean,max,sum]& \textbf{65.28 $\pm$ \, 6.33}&	55.00 $\pm$ 18.71&	\textbf{74.00 $\pm$ \, 7.35} & \textbf{63.67 $\pm$ \, 8.33}&	57.86 $\pm$ 13.96\\
    & & Graph-measures& flatten& 58.33 $\pm$ 14.91&  60.00 $\pm$ 25.50&  57.00 $\pm$ 21.35 & 55.33 $\pm$ 12.58&  55.43 $\pm$ 17.08\\
    & & LDW-FC &  [mean,max,sum] & 55.83 $\pm$ \, 8.07&	60.00 $\pm$ 20.00&	51.00 $\pm$ 18.55 & 52.00 $\pm$ \, 7.48&	54.22 $\pm$ 13.23\\
    
    \hline
    \multirow{6}{*}{\vtop{\hbox{\strut Supervised}\hbox{\strut GCN-FCNN}}}&\multirow{3}{*}{\vtop{\hbox{\strut Pearson}\hbox{\strut ($\tau= $0.25)}}}&Raw-fMRI&flatten &  57.78 $\pm$ 20.14&	65.00 $\pm$ 25.50&	52.00 $\pm$ 33.26 & 57.22 $\pm$ 19.61&	58.74 $\pm$ 18.32\\
    & & Graph-measures&[mean,max,sum]&62.78 $\pm$ 16.63&	50.00 $\pm$ 31.62&	75.00 $\pm$ 23.24 & 51.67 $\pm$ 34.32&	49.86 $\pm$ 31.49\\
    & & Pearson-FC& flatten& 56.11 $\pm$ 10.30&	60.00 $\pm$ 33.91&	56.00 $\pm$ 38.78 & 48.89 $\pm$ 31.70&	49.64 $\pm$ 24.94\\
    \cline{2-9}
    & \multirow{3}{*}{\vtop{\hbox{\strut LDW}\hbox{\strut ($\tau= $0.4)}}}& Raw-fMRI&flatten &53.61 $\pm$ 15.31&	25.00 $\pm$ 15.81&	\textbf{80.00 $\pm$ 25.30} & 51.67 $\pm$ 40.96&	32.05 $\pm$ 21.68\\
    & & Graph-measures& [mean,max,sum]&48.61 $\pm$ \, 9.86&	45.00 $\pm$ 33.17&	53.00 $\pm$ 28.57 & 38.89 $\pm$ 22.22&	39.45 $\pm$ 22.75\\
    & & LDW-FC &flatten& \textbf{62.50 $\pm$ \, 9.54}&	\textbf{60.00 $\pm$ 25.50}&	63.00 $\pm$ 31.56 & \textbf{61.67 $\pm$ 10.00}&	\textbf{57.86 $\pm$ 13.96}\\
    
\hline \hline
\end{tabular}}
\end{table*}

\begin{table*}[!t]
\centering
\caption{Classification performance of proposed models using different readout strategies for transforming learned embeddings as inputs to FCNN classifiers. Network adjacency matrix: LDW($\tau=$0.4). Node feature: LDW-FC.}
\label{table-readout}
\setlength{\tabcolsep}{10pt}
\renewcommand{\arraystretch}{1.1}
\resizebox{1\linewidth}{!}{
\begin{tabular}{llccccc}
\hline \hline
    Classifier &Readout & Acc & Sen & Spe & Pre & F1\\
\hline
    \multirow{5}{*}{Unsupervised GAE-FCNN}& flatten	&\textbf{72.50 $\pm$ 10.77}&	\textbf{60.00 $\pm$ 20.00}&	\textbf{83.00 $\pm$ 15.36} & \textbf{80.00 $\pm$ 18.71}&	\textbf{65.14 $\pm$ 17.20}\\
    &mean&	32.22 $\pm$ 10.66&	40.00 $\pm$ 33.91&	27.00 $\pm$ 28.21 & 24.56 $\pm$ 14.81&	29.75 $\pm$ 20.40\\
    &max&	46.39 $\pm$ 11.64&	55.00 $\pm$ 24.49&	40.00 $\pm$ 27.57 & 45.56 $\pm$ 12.37&	47.45 $\pm$ 11.92\\
    &sum&	39.44 $\pm$ 16.70&	35.00 $\pm$ 12.25&	43.00 $\pm$ 21.35 & 37.33 $\pm$ 18.03&	35.87 $\pm$ 14.66\\
    &$[$mean,max,sum$]$&	43.89 $\pm$ 12.47&	45.00 $\pm$ 18.71& 43.00 $\pm$ 11.66 &  40.00 $\pm$ 12.25&	42.00 $\pm$ 14.35\\
    \cline{1-7}
    \multirow{5}{*}{Unsupervised VGAE-FCNN}& flatten	&55.56 $\pm$ 15.32&	65.00 $\pm$ 25.50&	46.00 $\pm$ 21.31 & 51.67 $\pm$ 15.28&	\textbf{56.43 $\pm$ 17.72}\\
    &mean&	50.56 $\pm$ 17.07&	45.00 $\pm$ 24.49&	\textbf{54.00 $\pm$ 34.41} & 39.67 $\pm$ 24.37&	41.89 $\pm$ 24.19\\
    &max&	55.83 $\pm$ 17.00&	60.00 $\pm$ 33.91&	52.00 $\pm$ 43.08 & \textbf{52.46 $\pm$ 33.69}&	51.55 $\pm$ 26.63\\
    &sum&	51.39 $\pm $ \, 9.86&	50.00 $\pm$ 22.36&	53.00 $\pm$ 26.00 & 55.00 $\pm$ 24.49&	47.00 $\pm$ 13.27\\
    &$[$mean,max,sum$]$& \textbf{55.83 $\pm$ \, 8.07}&	\textbf{60.00 $\pm$ 20.00}& 51.00 $\pm$ 18.55 &	52.00 $\pm$ 7.48&	54.22 $\pm$ 13.23\\
    \cline{1-7}
    \multirow{5}{*}{Supervised GCN-FCNN}& flatten	&\textbf{62.50 $\pm$ \, 9.54}&	60.00 $\pm$ 25.50&	63.00 $\pm$ 31.56 & \textbf{61.67 $\pm$ 10.00}&	\textbf{57.86 $\pm$ 13.96}\\
    &mean&	57.50 $\pm$ 13.82&	\textbf{65.00 $\pm$ 33.91}&	48.00 $\pm$ 41.18 & 45.57 $\pm$ 25.21&	52.58 $\pm$ 27.09\\
    &max&	50.83 $\pm$ \, 7.01&	25.00 $\pm$ 31.62& \textbf{73.00 $\pm$ 24.82} &	20.00 $\pm$ 24.49&	22.00 $\pm$ 27.13\\
    &sum&	56.39 $\pm $ 13.54&	50.00 $\pm$ 35.36&	60.00 $\pm$ 20.98 & 40.33 $\pm$  24.64&	44.33 $\pm$ 27.78\\
    &$[$mean,max,sum$]$&	44.44 $\pm$ \, 6.09&	50.00 $\pm$ 27.39&	40.00 $\pm$ \, 8.16 & 40.00 $\pm$ \, 8.16&	42.76 $\pm$ 14.38\\
\hline \hline

\hline
\end{tabular}}
\end{table*}



\subsubsection{Effect of Network Thresholding} To assess the effect of network thresholding on the subsequent functional network classification, we examined the proportional thresholding (PT) and an alternative approach based on the local k-nearest neighbor graph (k-NNG) of FC matrices. While the PT applies a global threshold to select a $\tau$-fraction of the strongest connections in each individual network, the local k-NNG approach applies a local threshold to FC matrix, selecting the $k$-strongest edges of each node in the network \cite{van2017proportional}. By preserving the same node degrees in each network, the k-NNG like the PT can produce consistent network densities across subjects, and thus enabling meaningful between-group comparison or discrimination of network topology. We also studied the impact of varying thresholding values on the classification performance.

Fig.~\ref{fig:PT_KNN} shows the classification accuracies of the unsupervised GAE-FCNN model using the two network thresholding strategies over a range of tested thresholds. For PT (Fig.~\ref{fig:PT_KNN}(a)), the highest accuracy was achieved at $\tau=0.4$, supporting our choice of this threshold on this dataset to produce optimal topological information in the constructed networks for discriminating MDD and HC. As expected, use of smaller thresholds ($\tau < 0.4$) degrades the performance, owing to the removal of informative connectivity edges. On the other hand, larger thresholds ($\tau > 0.4$) may introduce more spurious or weak connections in the networks, and hence decreased classification accuracy. For k-NNG (Fig.~\ref{fig:PT_KNN}(b)), the classification performance is lower than the PT with highest accuracy at $k=65$. This may be due to the inclusion of more edges corresponding to weak (and thus less reliable) correlations compared to the application of a global PT, since the k-NNG approach mandates a fixed number of edges for all nodes.

\begin{table*}[!ht]
\centering
\caption{Performance comparison of proposed GAE-FCNNs with various state-of-the-art methods for functional connectome-based classification of MDD and HC. All methods used LDW-estimated correlations in rs-fMRI as FC features and network adjacency matrices constructed with $\tau$= 0.4}.

\label{table-cmpr}
\setlength{\tabcolsep}{10pt}
\renewcommand{\arraystretch}{1.3}
\begin{tabular}{llccccc}
\hline \hline
    &Classifier & Acc & Sen & Spe & Pre & F1\\
\hline 
    \multirow{6}{*}{Competing}&SVM-RBF& 50.83 $\pm$ \, 7.01&	15.00 $\pm$ 20.00& 82.00 $\pm$ 22.27 &	16.67 $\pm$ 21.08&	15.71 $\pm$ 20.40\\
	&BrainNetCNN \cite{kawahara2017brainnetcnn}	&51.11 $\pm$ \, 4.16&	45.00 $\pm$ 36.74&	57.00 $\pm$ 35.72 & 28.57 $\pm$ 23.47&	34.91 $\pm$ 28.57\\    
    &Population-based GCN \cite{kipf2016semi}	&55.56 $\pm$ 11.65&	45.00 $\pm$ 18.71&	66.00 $\pm$ 26.34 & 58.57 $\pm$ 20.90&	47.58 $\pm$ 12.84\\
    & GroupINN \cite{yan2019groupinn}	& 56.11 $\pm$ 11.93&	46.67 $\pm$ 32.32&	68.00 $\pm$ 32.50 & 41.67 $\pm$ 33.33&	37.33 $\pm$ 21.33\\
    & Hi-GCN \cite{jiang2020hi}	& 58.61 $\pm$ 10.03 &	40.00 $\pm$ 22.61 &	72.00 $\pm$ 20.40 & 43.33 $\pm$ 35.51 &	39.05 $\pm$ 25.27\\
    & E-Hi-GCN \cite{li2021te}	& 51.67 $\pm$ 18.39 &	46.67 $\pm$ 45.22 &	76.00 $\pm$ 23.32 & 16.67 $\pm$ 13.94 &	23.71 $\pm$ 20.52\\
    \hline
    \multirow{3}{*}{Proposed}&Supervised GCN-FCNN& 62.50 $\pm$ \, 9.54&	60.00 $\pm$ 25.50&	63.00 $\pm$ 31.56 & 61.67 $\pm$ 10.00&	57.86 $\pm$ 13.96\\
    &Unsupervised GAE-FCNN &\textbf{72.50 $\pm$ 10.77} &\textbf{60.00 $\pm$ 20.00} &\textbf{83.00 $\pm$ 15.36} & \textbf{80.00 $\pm$ 18.71} &\textbf{65.14 $\pm$ 17.20}\\
    &Unsupervised VGAE-FCNN &65.28 $\pm$ \, 6.33 &55.00 $\pm$ 18.71 &74.00 $\pm$ \, 7.35 &63.67 $\pm$ \, 8.33 &57.86 $\pm$ 13.96\\
\hline \hline
\end{tabular}
\end{table*}

\begin{figure*}[!ht]
\centerline{\includegraphics[width=1.8\columnwidth]{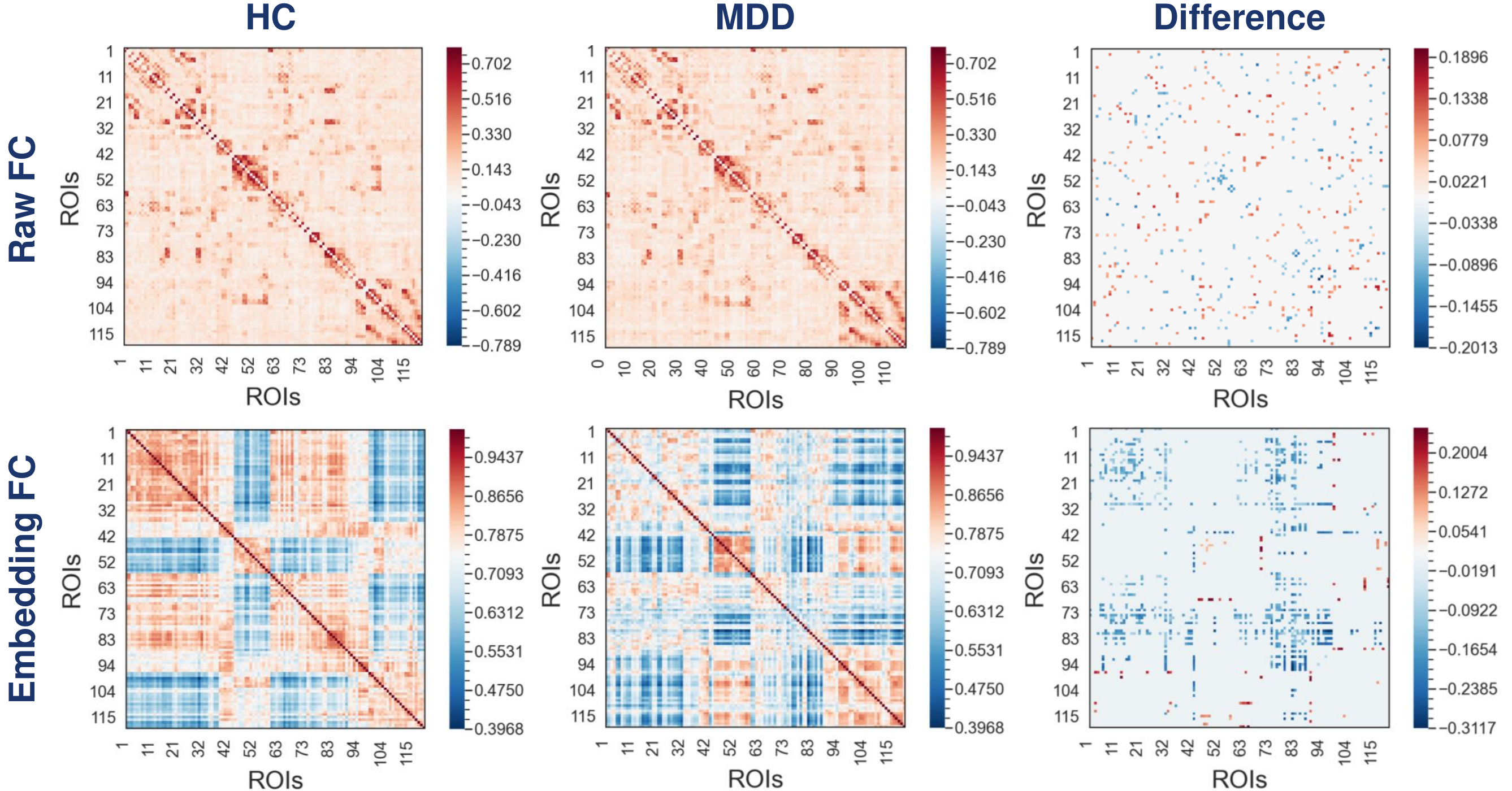}}
\caption{Differences in connectivity pattern between MDD and HC as revealed by raw FC from rs-fMRI data (Top) and high-order FC derived from GAE-learned node embeddings (Bottom). Group mean FC maps for HC (left) and MDD (middle) subjects, and their differences (MDD - HC) (right). The group differences shown are significantly different from zero at level $\alpha$ = 0.05 using an independent two-sample t-test.}
\label{fig:res_fig2}
\end{figure*}

\subsubsection{Comparison of Readout Strategies}

Table \ref{table-readout} shows the classification results for different readout strategies. We compared different readout/transformation methods to obtain graph-level representation $\mathbf{z}_G$ as inputs to FCNN classifier, i.e., flattening of $\mathbf{Z}$ and $\text{mean/max/sum}$ aggregation of node embeddings $\{\mathbf{z}_i\}$. It can be seen that the flattening method by concatenating learned embeddings of all nodes as input yields better classification performance for different classifiers generally, compared to the aggregation method which may induce loss of information about individual nodes.

\subsubsection{Comparison with State-of-the-Art Methods}

Table \ref{table-cmpr} shows the performance comparison of different connectome-based classification methods. The proposed methods clearly outperformed the competing models by a large margin, with the unsupervised GAE-FCNN performing the best. In consistency with recent studies, our results suggest the advantages of DNN methods over traditional SVM classifier with significant improvement in FC classification. The population-based GCNs perform slightly better than the BrainNetCNN. The population-based GCNs, while leveraging on pairwise associations between subjects in a population graph for node/subject-level classification, do not classify brain networks directly as in our proposed models. The use of grid-wise convolutions in the BrainNetCNN, despite its capability to capture spatial information of neighboring nodes, fails to account for irregular structure of brain networks. The superior performance of GAE-FCNN models compared to other DNNs implies that incorporating network topological structure as captured by the network embeddings for classification can provide discriminative information for identifying MDD, a disorder associated with disrupted brain networks. The proposed framework achieved the best accuracy of $72.5\%$ when using the unsupervised GAE-FCNN on a challenging task of classifying MDD brain networks, based on 5-fold CV on a small dataset, in contrast to the leave-one-subject-out classification in previous studies.

\subsubsection{Connectivity Maps Learned by GAE}


In Fig.~\ref{fig:res_fig1}, we plot the averaged feature maps of node-level embeddings learned by the GCN-GAE from LDW-based networks for the MDD and HC subjects. Noticeable difference in the learned embedding pattern can be seen between the two groups, with stronger activation for some ROIs in MDD compared to HC. Considerable between-variance is observed, indicating separability of the learned embeddings between two groups. This demonstrates the ability of the proposed model to extract latent representations of brain network structure that can clearly distinguish between MDD and controls, which explains the enhanced performance in the downstream classification task compared to other methods.

\begin{figure}[!t]
\centerline{\includegraphics[width=1\columnwidth]{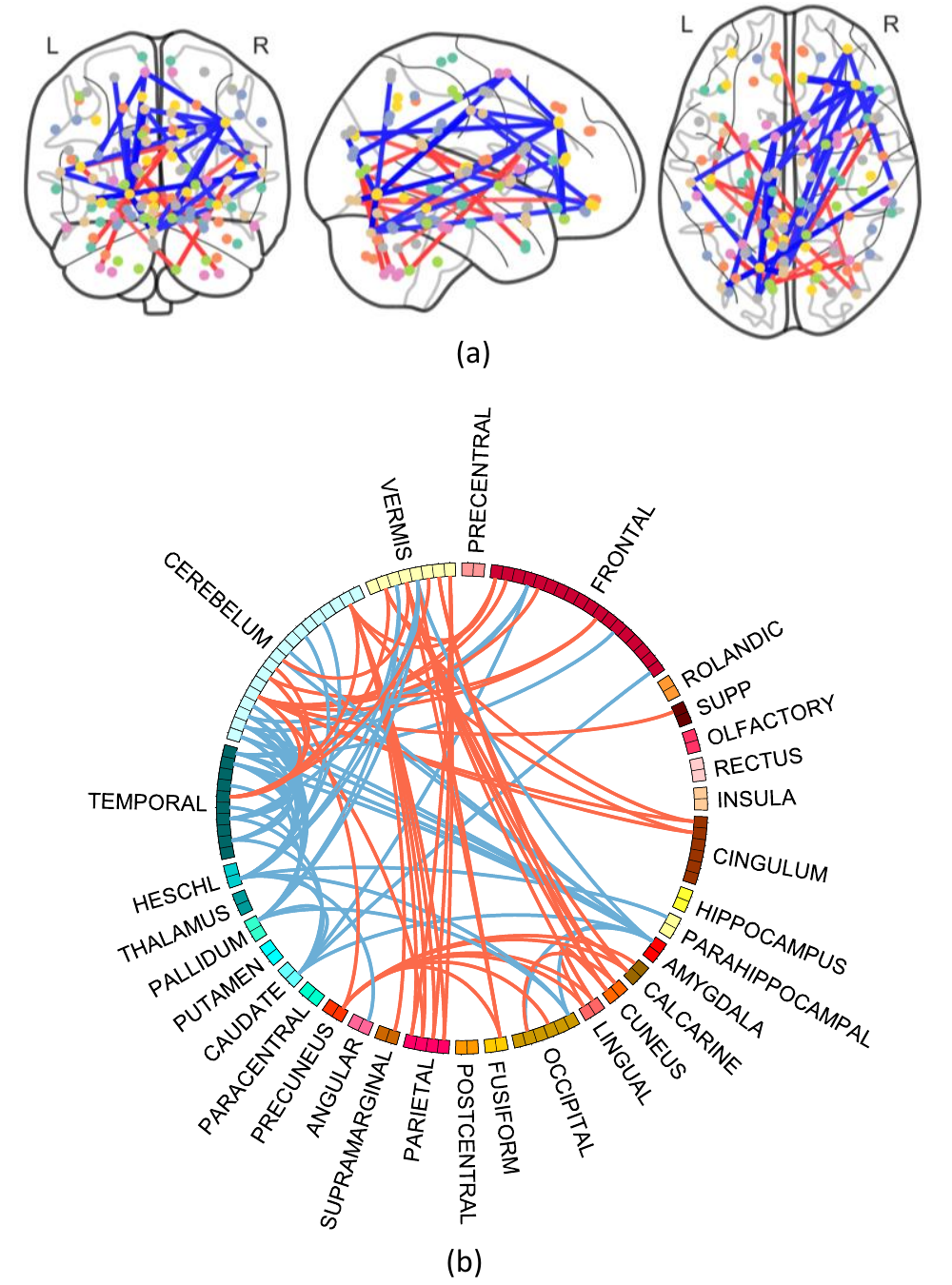}}
\caption{Difference in brain networks between HC and MDD detected by the high-order embedding-derived FC. (a) Topological representations show increase (red) and decrease (blue) in FC for MDD relative to HC. Only edges with significant differences in FC strength are shown ($p$\textless0.05). (b) Connectogram illustrates the corresponding changes in FC between modules of functionally-related brain regions. Top 100 connectivity edges with largest increase (red) and decrease (blue) in FC are shown.}
\label{fig:res_fig3}
\vspace{-0.1in}
\end{figure}

We further constructed high-order FC by correlating the GAE-learned embeddings $\mathbf{z}_{i}$ between pairs of nodes. Fig.~\ref{fig:res_fig2} shows the difference in connectivity pattern between the MDD and HC groups as quantified by the LDW-estimated raw FC and the embedding-based high-order FC. A group-level t-test was used to contrast the FC between the two groups, and connections with significant difference $(p$\textless0.05) are shown in Fig.~\ref{fig:res_fig2}(right). The embedding-based FC matrices (Fig.~\ref{fig:res_fig2}(left \& middle)) exhibit an apparent block structure revealing the modular organization, an important property of brain networks. Compared to raw FC, it is evident that the embedding-based FC detected more pronounced and systematic difference in connectivity, particularly between specific communities or modules of ROIs. To examine whether these differences are biologically meaningful and related to MDD as a network-based disorder, we plot the topological maps in Fig.~\ref{fig:res_fig3} to visualize the increase and decrease in FC between ROIs in MDD relative to HC. The embedding FC identified a spread reduction in intrinsic connectivity of the amygdala with a variety of ROIs involved in emotional processing and regulation in MDD subjects (including caudate, temporal regions, occipital cortex, and cerebellum), as reported in previous rs-fMRI studies \cite{ramasubbu2014}. In agreement with previous findings \cite{mulders2015}, we also found significant increase in FC in the default mode network (DMN). The detected altered rs-FC between cerebellum with the DMN and affective network has also been associated with major depression \cite{o2010,zeng2012}.

\section{Discussion}

We developed a deep GNN framework for embedding learning in brain functional networks to identify connectome-specific bio-signatures for classifying brain disorders such as MDD. The proposed GAE-FCNN provides a novel approach to incorporating the non-Euclidean information about graph structure into the classification of brain networks. It combines a GCN-based GAE that can learn latent embeddings effectively to encode topological information and node content, and a deep FCNN that leverages on the learned embeddings to reveal disrupted neural connectivity patterns in MDD relative to HC for classification purpose. On a challenging task of classifying MDD and HC using a small amount of rs-fMRI data, the proposed method substantially outperforms several state-of-the-art brain connectome classifiers, achieving the best accuracy of $72.5\%$ with the unsupervised GAE-FCNN model. Furthermore, high-order networks constructed from the node embeddings generated from the proposed GAE detects altered FC patterns in MDD related to emotional processing, which are not captured by the original FC measures. Our framework is generally applicable to other functional neuroimaging data, e.g., EEG-derived networks, and other neuropsychiatric disorders besides MDD associated with neural network dysconnectivity, showing potential as diagnostic tool in clinical settings.

There are potential limitations of our approach. First, our method focuses on embedding learning and classification for static brain networks. However, recent rs-fMRI studies suggest the temporal dynamicity of brain FC networks in which connectivity edges between regions evolve over time \cite{hutchison2013,samdin2016,ting2017}. Certain neuropsychiatric disorders have also been associated with disruptions in dynamic FC and graph properties such as in MDD \cite{zhi2018}. Future work could extend the proposed GAE framework to learn latent representations to embed the time-evolving network structure, by using some recent extensions of GCNs for dynamic graphs in the encoder part, e.g., the EvolveGCN \cite{pareja2020} which uses a recurrent neural network (RNN) to evolve the GCN parameters. Second, we analyzed a single type of brain networks from one neuroimaging modality, i.e., functional networks from fMRI. Multimodal fusion by combining different imaging modalities such as fMRI and diffusion imaging \cite{kang2017} could provide multiple views and hence more complete understanding of the brain networks. One possible direction is to characterize the fusion of functional and structural networks as multilayer networks, i.e., networks that can model multiple types of interactions and relations between brain nodes. Our GAE model can be generalized to produce embeddings for these multilayer brain networks, by incorporating the recently proposed multilayer GCN layers \cite{grassia2021} in the encoding phase. Moreover, our study uses a single type of node features for classification. One could explore different fusion strategies to learn embeddings for multiple node features in a unified way.
Third, our decoder model is designed to reconstruct the network structure only, which is adequate to learn embeddings to capture node relational information in brain networks. It could be extended to reconstruct both the input node features and the adjacency matrix to learn joint embeddings of both network structure and features to improve classification. This could be done by generalizing the decoder function (\ref{Eqn:dec}) as $p(\mathbf{A},\mathbf{X}|\mathbf{Z}) = p(\mathbf{A}|\mathbf{Z})p(\mathbf{X}|\mathbf{Z})$, and the reconstruction loss (\ref{Eqn:gaeL}) to $\mathcal{L} = \mathcal{L}_{\mathbf{A}} + \mathcal{L}_{\mathbf{X}}$ where $\mathcal{L}_{\mathbf{A}} = \mathbb{E}_{q(\mathbf{Z} \mid(\mathbf{X}, \mathbf{A}))}[\log p(\mathbf{A} \mid \mathbf{Z})]$ and $\mathcal{L}_{\mathbf{X}} = \mathbb{E}_{q(\mathbf{Z} \mid(\mathbf{X}, \mathbf{A}))}[\log p(\mathbf{X} \mid \mathbf{Z})]$. Finally, while this study has devised a novel framework producing network embeddings that differentiate MDD and HC and improve brain connectome classification, the interpretability of the model is important for clinical applications to understand the underlying mechanism behind the predictions and the neurobiological system being classified, instead of being used as a black box. Further studies could explore recent approaches to explaining the predictions in graph neural networks \cite{yuan2020}, e.g., to identify which input edges and node features of the brain networks are more important in predicting a certain disease class.


\vspace{-0.1in}
\appendix
This appendix contains details of hyper-parameter settings and additional analyses. We conducted additional performance evaluation on an independent MDD dataset to test the robustness of the proposed GAE-FCNN model and its generalization capability. We followed similar experimental setup of 5-fold CV, and we compared the performance with several state-of-the-art benchmark methods. Additionally, we investigated the performance when using different brain parcellation atlases.

\vspace{-0.1in}
\subsection{Hyper-parameter Settings}
Table~\ref{params} summarizes the list of hyper-parameter search space for the proposed and other methods used in experiments. 

\vspace{-0.15in}
\subsection{Additional Evaluations on Independent Dataset}

\subsubsection{Dataset and Preprocessing}
We used rs-fMRI data from the open-access REST-meta-MDD Consortium database \cite{yan2019reduced} for evaluation. We considered the largest dataset from site 20, consisting of a total 477 subjects (250 MDD and 227 HC) recruited from the Southwest University China. The rs-fMRI scans were acquired using Siemens scanner with an echo-planar imaging sequence (TR/TE = 2000/30 ms; flip angle = 90\textdegree; thickness/gap = 3.0/1.0 mm; time points = 242; field of view = 220 mm; voxel size = 3.44 × 3.44 × 4.00; matrix size = 61 × 73  × 61). The data were preprocessed using the Data Processing Assistant for Resting-State fMRI (DPARSF) \cite{yan2010dparsf}, following steps in \cite{yan2019reduced}.

\begin{table}[!ht]
\centering
\caption{Hyper-parameter search space of the proposed and compared methods.}
\label{params}
\resizebox{1\linewidth}{!}{
\begin{tabular}{lll}
\hline \hline
    Method	& Hyper-parameter &	Range \\
		\hline
    \multirow{3}{*}{SVM} & 	Kernel	& [Linear, RBF]\\
	& Regularization C &	$2^K$, $K$ = Discrete-uniform (min=-5, max=14, step=1)\\
	& Gamma	& $2^K$, $K$ = Discrete-uniform (min=-5, max=14, step=1)\\
	\hline
	\multirow{5}{*}{BrainNetCNN}	& Batch size	& Discrete-uniform (min=5, max=16, step=1)\\
	& Learning rate (LR)	& Log-uniform (min=1e-5, max=1e-2) \\
	& LR momentum 	& Log-uniform (min=1e-7, max=1e-2)\\
	& LR weight decay	& Discrete-uniform (min=0.1, max=0.9, step=0.1)\\
	& Others	& Default \\
	\hline
	\multirow{7}{*}{Population GCN} & Number of layers &	Min=1, max=3\\
    & Hidden dimensions	& $2^K$, $K$ = Discrete-uniform (min=4, max=16, step=1)\\
    & Optimizer	& Adam ($\beta_1$=0.9, $\beta_2$=0.999, $\epsilon$=1e-8)\\
    & LR	& Log-uniform (min=1e-5, max=1e-2)\\
    & LR scheduler reduce factor	& Discrete-uniform (min=0.1, max=0.9, step=0.1)\\
    & L2 weight decay	& Log-uniform (min=1e-7, max=1e-2)\\
    \hline
    \multirow{12}{*}{GroupINN} &  Batch size	& Discrete-uniform (min=5, max=32, step=1)\\
        & Optimizer	& Adam ($\beta_1$=0.9, $\beta_2$=0.999, $\epsilon$=1e-8)\\
        & LR	& Log-uniform (min=1e-5, max=1e-2)\\
        & L2 weight decay	& Log-uniform (min=1e-7, max=1e-2)\\
        & Input/output dimensions	& Discrete-uniform (min=5, max=16, step=1)\\
        & Embedding dimensions	& $2^K$, $K$ = Discrete-uniform (min=3, max=6, step=1)\\
        & Negative penalty 	& Discrete-uniform (min=0.1, max=0.9, step=0.1)\\
        & Negative penalty reduce	& Discrete-uniform (min=0.1, max=0.9, step=0.1)\\
        & Negative variance penalty 	& Discrete-uniform (min=0.1, max=0.9, step=0.1)\\
        & Positive variance penalty	& Discrete-uniform (min=0.1, max=0.9, step=0.1)\\
        & Negative orthogonal penalty	& Discrete-uniform (min=0.1, max=0.9, step=0.1)\\
        & Positive orthogonal penalty	& Discrete-uniform (min=0.1, max=0.9, step=0.1)\\

    \hline
    \multirow{5}{*}{Hi-GCN} & Optimizer	& Adam ($\beta_1$=0.9, $\beta_2$=0.999, $\epsilon$=1e-8)\\
        & fGCN LR	& Log-uniform (min=1e-5, max=1e-2)\\
        & pGCN LR	& Log-uniform (min=1e-5, max=1e-2)\\
        & pGCN output dimensions	& $2^K$, $K$ = Discrete-uniform (min=5, max=8, step=1)\\
        & Fully connected layer	& $2^K$, $K$ = Discrete-uniform (min=4, max=6, step=1)\\

    \hline
    \multirow{3}{*}{E-Hi-GCN} & Optimizer	& Adam ($\beta_1$=0.9  $\beta_2$=0.999, $\epsilon$=1e-8)\\
        & LR	& Log-uniform (min=1e-5, max=1e-2)\\
        & Input/output dimensions	& Discrete-uniform (min=5, max=16, step=1)\\

    \hline
	\multirow{15}{*}{\vtop{\hbox{\strut Unsupervised}\hbox{\strut V/GAE-FCNN}}} 	& Number of GCN layers	& Discrete-uniform (min=1, max=5, step=1)\\
	& GCN hidden dimensions	& $2^K$, $K$ = Discrete-uniform (min=4, max=8, step=1)\\
	& Batch size	& Discrete-uniform (min=5, max=32, step=1)\\
	& Optimizer	& Adam ($\beta_1$=0.9, $\beta_2$=0.999, $\epsilon$=0.001)\\
	& Learning rate (LR)	& Log-uniform (min=1e-5, max=1e-2)\\
	& LR scheduler reduce factor	& Discrete-uniform (min=0.1, max=0.9, step=0.1)\\
	& Aggregation	& [flatten, mean, max, sum] \\
	\cline{2-3}
	& Number of FCNN layers	& Discrete-uniform (min=1, max=5, step=1)\\
	& FCNN hidden dimensions	& $2^K$, $K$ = Discrete-uniform (min=5, max=8, step=1)\\
	& Batch size	& Discrete-uniform (min=5, max=32, step=1)\\
	& Optimizer	& Adam ($\beta_1$=0.9, $\beta_2$=0.999, $\epsilon$=1e-8)\\
	& LR	& Log-uniform (min=1e-5, max=1e-2)\\
	& L2 regularization	& Log-uniform (min=1e-5, max=1e-2)\\
	& Dropout 	& Discrete-uniform (min=0, max=0.5, step=0.1)\\
	& LR scheduler reduce factor	& Discrete-uniform (min=0.1, max=0.9, step=0.1)\\
	\hline
	\multirow{11}{*}{\vtop{\hbox{\strut Supervised}\hbox{\strut GCN-FCNN}}}	& Number of GCN layers	& Discrete-uniform (min=1, max=3, step=1)\\
	& GCN hidden dimensions	& $2^K$, $K$ = Discrete-uniform (min=4, max=8, step=1)\\
	& Batch size	& Discrete-uniform (min=5, max=32, step=1)\\
	& Optimizer	& Adam ($\beta_1$=0.9, $\beta_2$=0.999, $\epsilon$=1e-8)\\
	& LR	& Log-uniform (min=1e-5, max=1e-2)\\
	& L2 weight decay	& Log-uniform (min=1e-7, max=1e-2)\\
	& LR scheduler reduce factor	& Discrete-uniform (min=0.1, max=0.9, step=0.1)\\
	& Dropout 	& Discrete-uniform (min=0, max=0.5, step=0.1)\\
	& Aggregation	& [flatten, mean, max, sum] \\
	\cline{2-3}
	& Number of FCNN layers	& Discrete-uniform (min=1, max=5, step=1)\\
	& FCNN hidden dimensions	& $2^K$, $K$ = Discrete-uniform (min=5, max=8, step=1)\\

    \hline\hline
\end{tabular}}
\vspace{-0.15in}
\end{table}


\begin{figure*}[!t]
\centerline{\includegraphics[width=1.8\columnwidth]{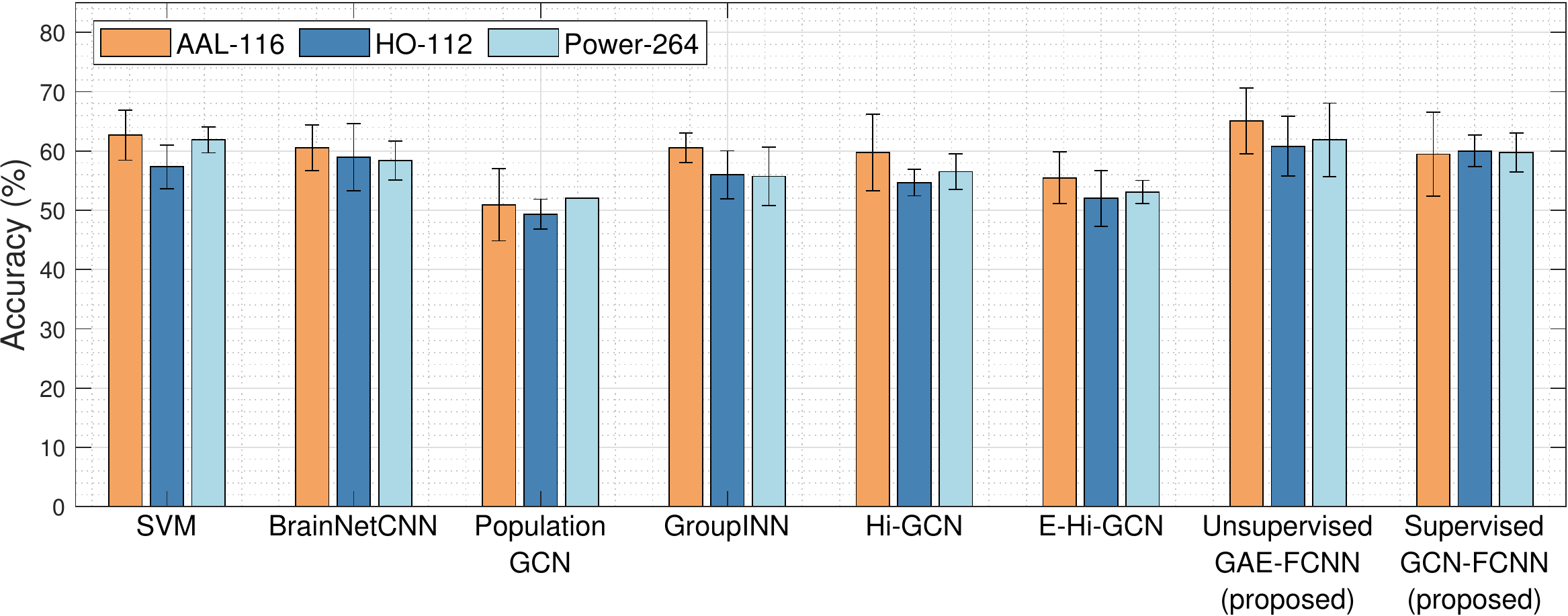}}
\caption{Performance comparison of using different brain parcellation atlases for rs-fMRI functional connectome-based classification of MDD and HC on REST-meta-MDD rs-fMRI dataset. Error bars indicate standard deviations of accuracy across partitioning folds in 5-fold cross-validation.}
\label{fig:atlases_all}
\vspace{-0.15in}
\end{figure*}

\begin{table*}[!t]
\centering
\caption{Performance comparison of proposed GAE-FCNN with various state-of-the-art methods for functional connectome-based classification of MDD and HC on REST-meta-MDD rs-fMRI dataset. All methods used LDW-estimated correlations in rs-fMRI as FC features and network adjacency matrices.}
\label{table-cmpr_rest}
\setlength{\tabcolsep}{10pt}
\renewcommand{\arraystretch}{1.3}
\begin{tabular}{llccccc}
\hline \hline
    &Classifier & Acc & Sen & Spe & Pre & F1\\
\hline 
    \multirow{6}{*}{Competing}&SVM-RBF&  62.67 $\pm$ 4.22 &	69.74 $\pm$ \, 5.48 &	55.00 $\pm$ \, 4.08 & 62.63 $\pm$ \, 3.39 &	65.97 $\pm$ \, 4.21\\
	&BrainNetCNN \cite{kawahara2017brainnetcnn}	& 60.53 $\pm$ 3.83 &	55.90 $\pm$ \, 5.94 &	\textbf{65.56 $\pm$ \, 7.97} & 64.14 $\pm$ \, 5.67 &	59.48 $\pm$ \, 4.21\\  
    &Population-based GCN \cite{kipf2016semi}& 50.93 $\pm$ 6.10 &	\textbf{80.00 $\pm$ 29.83} &	19.44 $\pm$ 25.82 & 50.29 $\pm$ \, 5.84 &	59.68 $\pm$ 16.41 \\
    & GroupINN \cite{yan2019groupinn}	&  60.53 $\pm$ 2.47 &	62.00 $\pm$ \, 2.67 & 57.22 $\pm$ \, 9.40 & 63.59 $\pm$ 11.05 &	62.15 $\pm$ \, 5.24\\
    & Hi-GCN \cite{jiang2020hi}	& 59.73 $\pm$ 6.44 &	61.17 $\pm$ \, 5.52 &	60.00 $\pm$ \, 3.77 & 59.49 $\pm$ 11.85 &	60.08 $\pm$ \, 8.65\\
    & E-Hi-GCN \cite{li2021te}	& 55.47 $\pm$ 4.35 &	58.62 $\pm$ \, 5.26 &	59.44 $\pm$ 13.45 & 51.79 $\pm$ 14.27 &	53.91 $\pm$ \, 7.17\\
    \hline
    \multirow{3}{*}{Proposed}&Supervised GCN-FCNN& 59.47 $\pm$ 7.09 &	54.87 $\pm$ \, 6.20 &	64.44 $\pm$ 12.35 & 63.66 $\pm$ \, 9.92 &	58.57 $\pm$ \, 6.02 \\
    &Unsupervised GAE-FCNN & \textbf{65.07 $\pm$ 5.56} &	69.74 $\pm$ \, 9.09 &	60.00 $\pm$ \, 7.16 & \textbf{65.38 $\pm$ \, 5.04} &	6\textbf{7.29 $\pm$ \, 6.22}\\ 
    &Unsupervised VGAE-FCNN & 60.79 $\pm$ 4.84 &	62.64 $\pm$ \, 5.34 &	58.78 $\pm$ \, 7.22 & 62.55 $\pm$ \, 5.00 &	62.50 $\pm$ \, 4.52\\
\hline \hline
\end{tabular}
\vspace{-0.05in}
\end{table*}

\subsubsection{Comparison of Different Brain Parcellations}
To examine the effect of different brain parcellations on the FC classification performance, we evaluated our method on the ROI-wise fMRI time series data extracted based on three parcellation atlases (both anatomical and functional): AAL atlas, Harvard-Oxford (HO) atlas (derived from anatomical landmarks: sulci and gyral) \cite{kennedy1998gyri}, and Power atlas (comprising functional areas associated with 13 large-scale functional networks and a group of unlabeled regions) \cite{power2011functional}, with respective number of ROIs of 116, 112 and 264. Fig.~\ref{fig:atlases_all} shows the MDD classification accuracies of different methods on the various brain atlases. It is apparent that the proposed supervised GCN-FCNN performs the best compared to other competing methods on all atlases. Among the atlases, AAL-116 generally gives better classification than HO and Power atlases over all methods. This suggests that FC networks based on anatomical ROIs may provide more discriminative information for differentiating between MDD and HC, compared to functional-ROI networks.

\subsubsection{Comparison with State-of-the-Art Methods}
Table~\ref{table-cmpr_rest} shows the comparison of our methods with several state-of-the-art brain FC classifiers under different classification performance metrics on the AAL-116 data. The selected architecture of the unsupervised GAE-FCNN is: two-layered GCN with embedding dimensions of 256 and 32, and three-layered FCNN with hidden dimensions of 256, 128 and 64. The proposed unsupervised GAE-FCNN achieved the best performance with average Acc of 65.07\% and F1 scores of 67.29\%, significantly outperforming the traditional SVM method, BrainNetCNN and other recent GCN models for brain network classification. The high sensitivity for population GCN with poor performance in other metrics is due to classification of all samples into one class. These results again attest to the superior performance of our method generalizable to other MDD fMRI dataset, suggesting the advantages of GAE-learned embeddings of network topology for brain connectomic classificaiton.


\bibliographystyle{IEEEbib}
\bibliography{References}

\end{document}